\documentclass[a4paper,twocolumn,showpacs,superscriptaddress,prl]{revtex4}
\usepackage[english]{babel}
\usepackage{graphicx}
\usepackage{xspace}

\begin{document}

\title{Possible pseudogap behavior of electron doped
high-temperature superconductors}

\author{S. Kleefisch}
\affiliation{II.~Physikalisches Institut, Universit\"{a}t zu K\"{o}ln,
Z\"{u}lpicher Str.~77, 50937 K\"{o}ln, Germany}

\author{B. Welter}
\affiliation{II.~Physikalisches Institut, Universit\"{a}t zu K\"{o}ln,
Z\"{u}lpicher Str.~77, 50937 K\"{o}ln, Germany}
\affiliation{Walther-Meissner-Institut, Bayerische Akademie
der Wissenschaften, Walther-Meissner Str.~8, 85748 Garching,
Germany}

\author{M. Naito}
\affiliation{NTT Basic Research Laboratories, 3-1 Morinosato Wakamiya,
Atsugi-shi, Kanagawa 243, Japan}

\author{A. Marx}
\affiliation{II.~Physikalisches Institut, Universit\"{a}t zu K\"{o}ln,
Z\"{u}lpicher Str.~77, 50937 K\"{o}ln, Germany}
\affiliation{Walther-Meissner-Institut, Bayerische Akademie
der Wissenschaften, Walther-Meissner Str.~8, 85748 Garching,
Germany}

\author{L. Alff}
\email{alff@wmi.badw.de}
\affiliation{II.~Physikalisches Institut, Universit\"{a}t zu K\"{o}ln,
Z\"{u}lpicher Str.~77, 50937 K\"{o}ln, Germany}
\affiliation{Walther-Meissner-Institut, Bayerische Akademie
der Wissenschaften, Walther-Meissner Str.~8, 85748 Garching,
Germany}

\author{R. Gross}
\affiliation{II.~Physikalisches Institut, Universit\"{a}t zu K\"{o}ln,
Z\"{u}lpicher Str.~77, 50937 K\"{o}ln, Germany}
\affiliation{Walther-Meissner-Institut, Bayerische Akademie
der Wissenschaften, Walther-Meissner Str.~8, 85748 Garching,
Germany}

\date{received October  09, 2000}

\begin{abstract}

We have measured the low-energy quasiparticle excitation spectrum
of the electron doped high-temperature superconductors (HTS)
Nd$_{1.85}$Ce$_{0.15}$CuO$_{4-y}$ and
Pr$_{1.85}$Ce$_{0.15}$CuO$_{4-y}$ as a function of temperature
and applied magnetic field using tunneling spectroscopy. At zero
magnetic field, for these optimum doped samples no excitation gap
is observed in the tunneling spectra above the transition
temperature $T_c$. In contrast, below $T_c$ for applied magnetic fields well
above the resistively determined upper critical field, a clear
excitation gap at the Fermi level is found which is comparable to
the superconducting energy gap below $T_c$. Possible
interpretations of this observation are the existence of a normal
state pseudogap in the electron doped HTS or the existence of a
spatially non-uniform superconducting state.

\end{abstract}


\maketitle

The existence of a pseudogap in hole doped high-temperature
superconductors (HTS) has been established over the recent years.
The physical origin of the pseudogap state, however, is still one
of the most debated topics for HTS. For a recent experimental
review see e.~g. \cite{Timusk:99}. In different types of
experiments including tunneling spectroscopy it has been found
that the pseudogap feature and the superconducting energy gap
merge smoothly into each other at the critical temperature $T_c$
\cite{Ding:96,Loeser:96,Renner:98,deWilde:98,Miyakawa:98}. Even
more, from angle-resolved photoemission experiments it has been
suggested that the pseudogap has the same $d_{x^2-y^2}$-symmetry
as the superconducting gap in the hole doped HTS
\cite{Ding:96,Loeser:96}. It has also been observed that the
temperature $T^{\star}$ associated with the appearance of the
pseudogap state roughly becomes equal to $T_c$ around optimum
doping or in the slightly overdoped regime, but is considerably
larger than $T_c$ in the underdoped regime. The evident question
arising from these experimental observations is whether or not
there is a relation between the physical origin of the
superconducting gap and the pseudogap. Such a scenario has been
proposed within theories involving so-called preformed pairs or at
least dynamical pair correlations above $T_c$ \cite{Emery:97}.

With respect to the different HTS materials, the hole doped system
La$_{2-x}$Sr$_x$CuO$_4$ seems to be a special case. For this
material, the behavior of the pseudogap has been reported to be
different compared to the other hole doped HTS, e.~g.~the size of
the pseudogap might be much larger than the superconducting gap
\cite{Sato:99}. However, the experimental situation is not well
settled and more experiments are needed to further clarify this
point. For the electron doped HTS of the class
$Ln_{2-x}$Ce$_x$CuO$_4$ ($Ln=\text{Nd, Pr}$) with $T'$ structure,
up to now no low-energy spectroscopic experiments probing the
pseudogap state have been reported. There is no doubt that
experiments on electron doped HTS are important and highly
desired with regard to the question whether hole and electron
doped HTS have the same underlying mechanism of superconductivity
and the pseudogap state. Furthermore, controversial experimental
results on the symmetry of the superconducting order parameter in
the electron doped HTS have been published recently
\cite{Kashiwaya:98,Alff:98,Alff:99,Tsuei:00}. That is, both the
symmetry of the order parameter and the question whether there is
a normal state pseudogap are under discussion for electron doped
HTS.

In this Letter, we report on the measurement of the tunneling
spectra in superconductor - insulator - superconductor junctions
based on  bicrystal grain boundary junctions (GBJs). The
temperature and magnetic field dependence of the tunneling
spectra has been studied up to 16\,T for the optimum electron
doped HTS Nd$_{1.85}$Ce$_{0.15}$CuO$_{4-y}$ (NCCO) and
Pr$_{1.85}$Ce$_{0.15}$CuO$_{4-y}$ (PCCO). While above $T_c(B=0)$ no
pseudogap feature could be observed, below $T_c(B=0)$ a pseudogap
around the Fermi level is clearly present for magnetic fields
larger than the resistively determined critical field
$B_{c2}^{\rho}$. This suggests that similar to the hole doped
HTS, there is a pseudogap state also for the electron doped HTS.
However, the presence of a non-uniform superconducting state may
also be consistent with our observations.

The NCCO- and PCCO-GBJs have been fabricated by the deposition of
200\,nm thick, $c$-axis oriented NCCO- and PCCO-films on
SrTiO$_3$ bicrystal substrates using molecular beam epitaxy
(MBE). Only symmetric, [001] tilt bicrystals with  misorientation
angles of $24^{\circ}$ and $36^{\circ}$ have been used. The thin
film deposition has been described in detail by Naito {\em et
al.} \cite{Naito:95}. For bicrystal GBJs with a junction area of
about 10$^{-8}$\,cm$^2$ a normal resistance ranging between 0.1
and 5\,k$\Omega$ is obtained \cite{Kleefisch:98}. Josephson
behavior in NCCO-GBJs has been demonstrated recently by Kleefisch
{\em et al.} \cite{Kleefisch:98}. We stress that the observation
of Josephson behavior for both NCCO and PCCO together with the
low in-plane resistivity ($\rho_{ab}<50\,\mu\Omega$cm at 25\,K)
of the films, $T_c$ values of about 24\,K, and the X-ray data
demonstrate that our thin film samples are optimum doped, well
oxygen reduced, and single phase. This is important with respect
to the possibility of an inhomogeneous dopant or oxygen
distribution, or the formation of different phases. We note that
such inhomogeneities are more difficult to be ruled out for large
bulk single crystals.


\begin{figure}[bt]
 \center{\includegraphics [width=1.0\columnwidth]{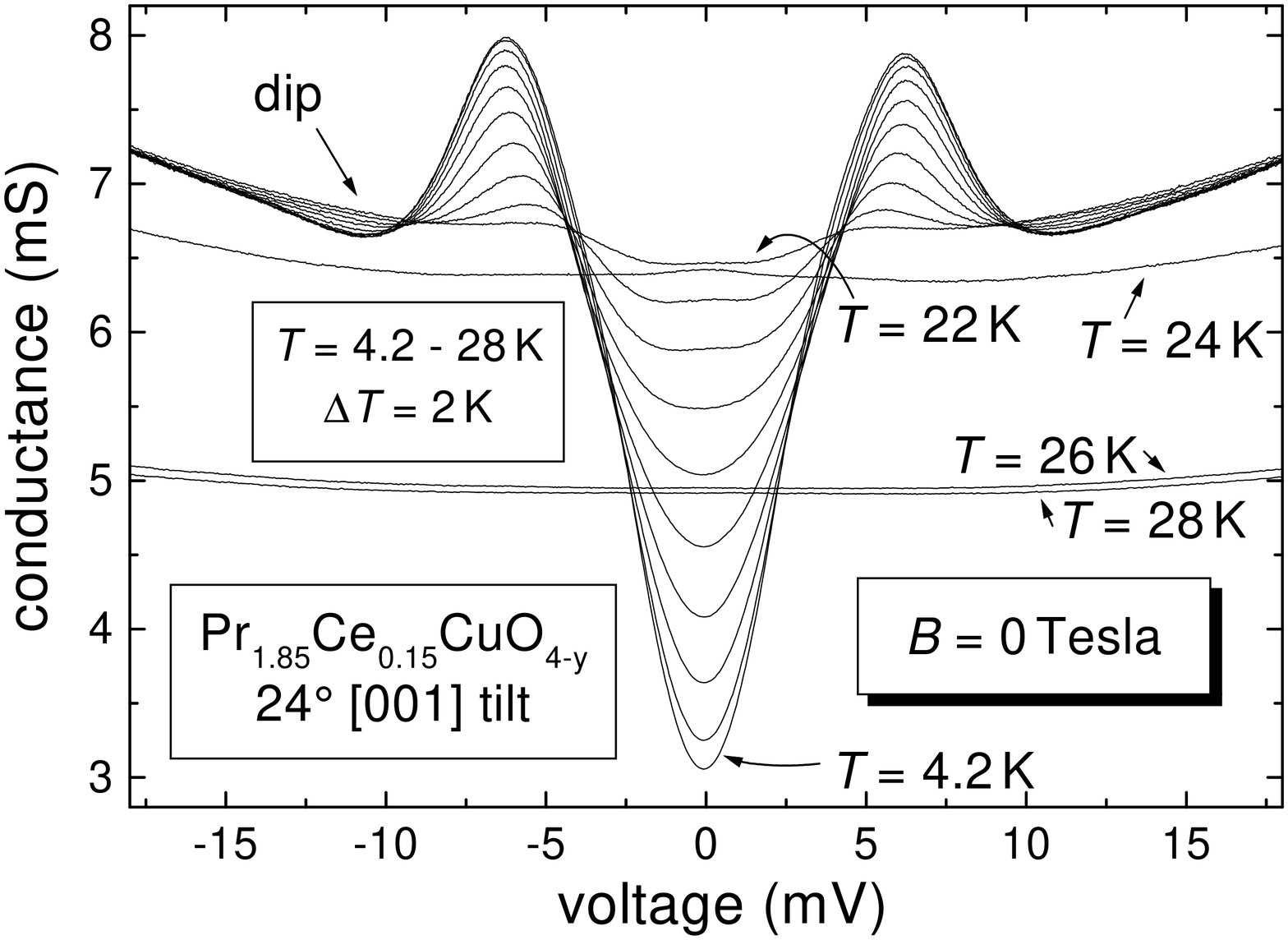}}
\vspace*{-0.2cm}
 \caption{Conductance vs. voltage of a $24^\circ$ [001] tilt
PCCO-GBJ in zero applied magnetic field measured at different
temperatures.}
 \label{tdep}
\end{figure}

In Fig.~\ref{tdep} quasiparticle tunneling spectra of a PCCO-GBJ
are shown for temperatures between $4.2$ and 28\,K  in zero
magnetic field. Very similar spectra are obtained for NCCO. The
superconducting gap clearly shows up in the spectra resulting in
a large density of states peak at $2\Delta$ (6.2\,meV for PCCO
and 5.8\,meV for NCCO at 4.2\,K). We note that the peak value
itself corresponds to $2\Delta$ only for a perfectly isotropic
$s$-wave symmetry of the superconducting order parameter. Fitting
the spectra to more anisotropic gap structures, larger values for
$\Delta$ up to 4\,meV are obtained depending on the degree of
anisotropy. Around $V=\pm10$\,mV a dip structure is observed in
the spectra that also has been seen in $c$-axis tunneling in
Bi$_2$Sr$_2$CaCu$_2$O$_{8+\delta}$ \cite{deWilde:98}. So far it
is not clear whether this dip structure indicates strong coupling
of the quasiparticles to collective excitations (as e.g.
phonons). Fig.~\ref{tdep} shows that the states in the
superconducting gap are filled up with increasing temperature.
While for NCCO almost no change of the peak position is observed
with increasing temperature, for PCCO there is a slight
reduction. However, this reduction is much smaller than expected
for $\Delta(T)$ according to BCS-theory. The gap structure
disappears at 24\,K which corresponds to the $T_c$ of the sample.
Fig.~\ref{tdep} also shows a temperature independent background
conductance of the GBJ that is parabolic in the low-energy
regime, becomes linear for voltages above 20-30\,mV, and stays
about linear up to several 100\,mV (not shown in Fig.~\ref{tdep}).


\begin{figure}[tb]
\center{\includegraphics [width=1.0\columnwidth]{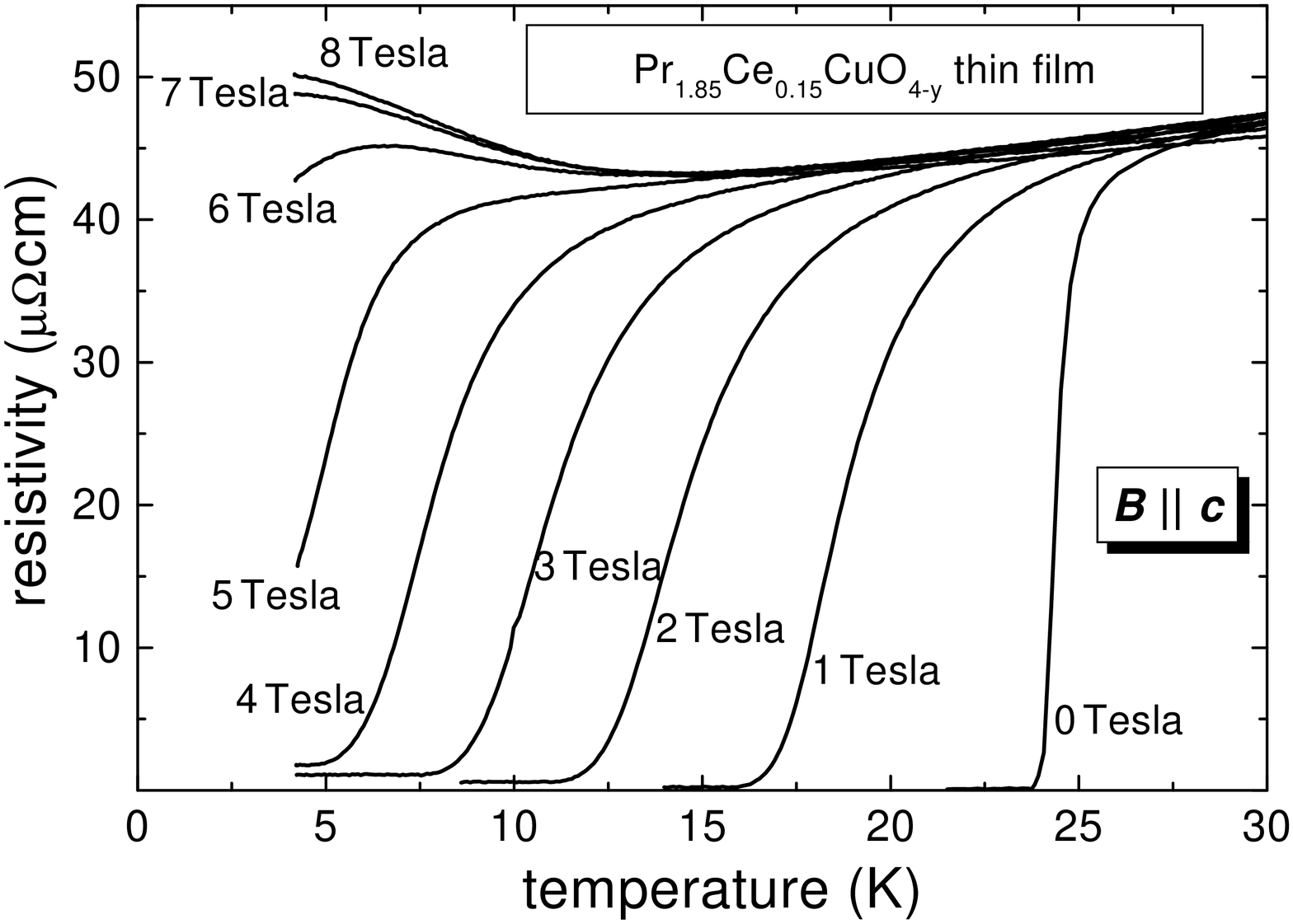}}
\vspace*{-0.2cm}
 \caption{Resistive transition of a  Pr$_{1.85}$Ce$_{0.15}$CuO$_{4-y}$
thin film sample at different applied magnetic fields.}
 \label{rtb}
\end{figure}

Another way to probe the normal state properties of a
superconducting sample is to apply a magnetic field that is
larger than its upper critical field. Following the experimental
approach of \cite{MacKenzie:93,Gollnik:98}, we have measured the
magnetic field dependence of the resistive transition of a PCCO
epitaxial thin film with the magnetic field applied parallel to
the $c$-axis. As shown in Fig.~\ref{rtb}, in a magnetic field of
1\,T, the superconducting onset temperature is reduced and the
transition width is increased. On further increasing the applied
field, $\rho_{ab}(B,T)$ further shifts to lower temperatures,
however, no further broadening of the superconducting transition
is observed as it is the case e.~g.~for
YBa$_2$Cu$_3$O$_{7-\delta}$. From the data in Fig.~\ref{rtb} one
can derive an upper critical field which is referred to as the
resistive critical field $B_{c2}^{\rho}(T)$. The size of
$B_{c2}^{\rho}(T)$ depends on the chosen resistivity criterion.
The functional form of $B_{c2}^{\rho}(T)$, however, seems to
remain the same within our experimental resolution (see discussion below).


\begin{figure}[tb]
\center{\includegraphics [width=1.0\columnwidth]{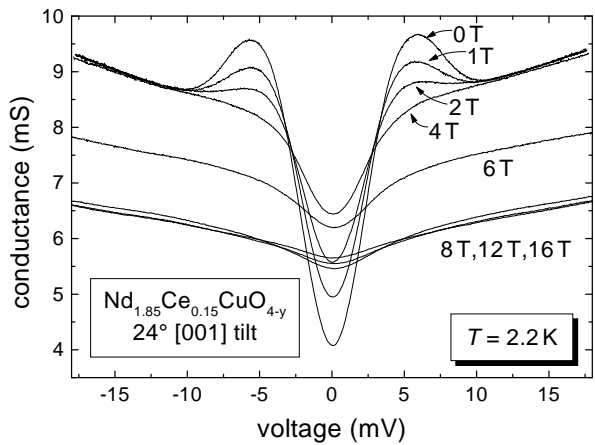}}
\vspace*{-0.2cm}
 \caption{Conductance vs. voltage curves of a symmetric $24^\circ$ [001] tilt
NCCO-GBJ measured at 2.2\,K in different applied magnetic fields applied
parallel to the $c$-axis.}
 \label{bdep}
\end{figure}

Fig.~\ref{bdep} shows the quasiparticle tunneling spectra for NCCO
measured at 2.2\,K for magnetic fields between 0 and 16\,T applied
parallel to the $c$-axis.  The main effect of the applied field is
the suppression of the density of states peaks at the
superconducting gap feature and the filling of the gap at smaller
voltages. Note that the position of the peaks does not change
with varying applied magnetic field, however, the peak amplitude
decreases with increasing field and disappears at
$B_{c2}^{\rho}$, which is about 5.6\,T at 2.2\,K. A key
experimental finding is the fact that the gap feature itself
remains clearly present even for the largest applied field of
16\,T. This shows the {\em existence of a gapped state for
$B>B_{c2}^{\rho}$}.

\begin{figure}[bt]
\center{\includegraphics [width=1.0\columnwidth]{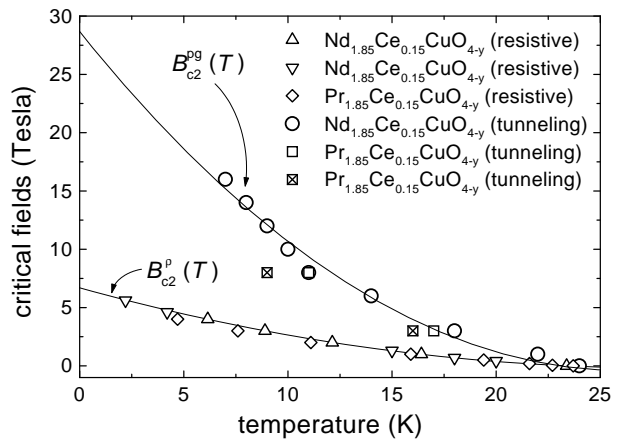}}
\vspace*{-0.2cm}
 \caption{Temperature dependence of the resistive
critical field $B_{c2}^{\rho}$ (full symbols) and the pseudogap critical
field $B_{c2}^{\text{pg}}$ (open symbols) for different samples of the
electron doped HTS NCCO and PCCO. The solid lines are guides to the eye.}
 \label{bc2}
\end{figure}

While at 2.2\,K the magnetic field needed to close the pseudogap
feature is beyond the maximum field of 16\,T available in our
experiments, this is not the case for $T>7$\,K. In this $T$
regime, we can define a pseudogap critical field
$B_{c2}^{\text{pg}}$ that is sufficient to close the pseudogap
feature. In Fig.~\ref{bc2} we show the temperature dependence of
both $B_{c2}^{\rho}$ and $B_{c2}^{\text{pg}}$. Although there is
significant scatter in our data due to broadening, it is evident
that $B_{c2}^{\text{pg}}$ is by a factor of about 4 larger than
$B_{c2}^{\rho}$. This observation holds for both NCCO and PCCO
indicating that there is no major difference among these electron
doped materials arising e.g. from the magnetic moments of the
Nd$^{3+}$ ions in NCCO \cite{Alff:99}. It is interesting to note
that the functional form of the temperature dependence of both
critical fields deviates strongly from a BCS-type temperature
dependence of the upper critical field having negative curvature.
The unusual positive curvature of $B_{c2}^{\rho}(T)$ in
Fig.~\ref{bc2} for NCCO has been also reported by
\cite{Hidaka:89,Gollnik:98}, and, furthermore, has been observed
in the hole doped HTS Tl$_2$Ba$_2$CuO$_{6+\delta}$
\cite{MacKenzie:93}. Recent measurements of Bi$_2$Sr$_2$CuO$_x$
have shown that the curvature depends on the chosen resistivity
criterion, becoming almost linear for a high resistivity
criterion as expected for conventional type-II superconductors
\cite{Vedeneev:99}. For our case, the curvature seems not to
depend on the chosen criterion, however, for large fields and
high resistivity criteria the evaluation of $B_{c2}^{\rho}(T)$
becomes difficult due to the broadened $\rho(T)$.

While for the unusual behavior of $B_{c2}^{\rho}(T)$
intrinsic origins have been proposed in the context of a quantum
critical point at $T=0$ \cite{Kotliar:96} and also within a
bipolaron theory \cite{Alexandrov:97}, recently Geshkenbein {\em
et al.}~have suggested that inhomogenous superconducting
properties (as for example due to an inhomogeneous oxygen or
dopant distribution) can cause the observed positive curvature of
$B_{c2}^{\rho}(T)$ \cite{Geshkenbein:98}. In this case one has to
assume regions with increased local $T_c$ compared to the bulk
$T_c$. Since in our case we are dealing with homogeneously
reduced thin films close to optimum doping level, an explanation
based on inhomogeneous oxygen or dopant distribution is neither
likely nor expected to change $T_c$ considerably. However,
intrinsically non-uniform superconductivity due to e.g.~phase
separation cannot be ruled out. Any further discussion of these
issues is beyond the scope of this Letter.


We now address possible origins of the observed quasiparticle
excitation gap structure observed below $T_c$ for
$B_{c2}^{\rho}(T)<B<B_{c2}^{\text{pg}}(T)$. Of course, it is
tempting to assume that for electron doped cuprates there is a
pseudogap feature with similar properties as has been observed
for the hole doped HTS. Then, according to our data the pseudogap
in the electron doped HTS merges smoothly into the
superconducting energy gap at $B_{c2}^{\rho}$ in analogy to the
experimental observation in
\cite{Renner:98,deWilde:98,Miyakawa:98} that the pseudogap in the
hole doped HTS merges into the superconducting gap at $T_c$.
Moreover, in both cases the density of states peak in the
tunneling spectra disappears at the transition from the
superconducting into the pseudogap state. We note that a recent
$c$-axis tunneling study on Bi$_2$Sr$_2$CaCu$_2$O$_{8+\delta}$
does not support the merging of the superconducting gap and the
pseudogap \cite{Krasnov:00}. A spatially non-uniform
superconducting state with regions having locally higher $B_{c2}$
than the bulk could also produce a gap like behavior in
quasiparticle tunneling. However, as discussed above it is
unclear how such regions can form within the optimum electron and
oxygen doped compound.

We next discuss possible physical interpretations of the critical
field $B_{c2}^{\text{pg}}$. Let us first assume that
$B_{c2}^{\text{pg}}$ is close to the thermodynamic critical field
$B_{c2}$. Then, extra\-polating $B_{c2}^{\text{pg}}(T)$ to $T=0$
yields $B_{c2}(T=0)\approx30$\,T. Using the relation
$\xi_{ab}(0)=\sqrt{\Phi_0/2\pi B_{c2}(0)}$, one derives
$\xi_{ab}(0) \approx 30$\,{\AA}. This value is similar to that
obtained e.~g.~for the hole doped HTS
La$_{1.85}$Sr$_{0.15}$CuO$_4$ but is significantly smaller than
the usual value of about 70\,{\AA} in NCCO \cite{Hidaka:89}. A
smaller value of $\xi_{ab}(0)$ would increase the importance of
fluctuation effects in the electron doped HTS \cite{Emery:97}.
However, as can be seen from Fig.~\ref{rtb}, in the resistivity
vs temperature curves the fluctuation regime is small and fields
slightly above $B_{c2}^{\rho}$ are sufficient to drive the films
onto a weakly field dependent semiconducting $\rho_{ab}(T)$ curve.
Moreover, the fluctuation analysis in \cite{Gollnik:98} suggests
that $B_{c2}^{\rho}$ is not identical with the thermodynamic
critical field $B_{c2}$, but cannot account for the difference by
a factor of around four between $B_{c2}^{\rho}$ and
$B_{c2}^{\text{pg}}$. Lastly, we note that recently a very small
value of $\xi_{ab}\approx 6-9$\,{\AA} has been calculated assuming a
spin-fluctuation pairing mechanism for both hole and electron
doped HTS \cite{Manske:00} corresponding to an even higher value
of $B_{c2}$.

Finally, at present there is no prediction for the magnetic field
dependence of the pseudogap even in the hole doped case. Assuming
that the pseudogap is caused by the existence of so-called
preformed pairs, the pseudogap state should be destroyed in
magnetic fields exceeding the Clogston paramagnetic limit
\cite{Clogston:62} $B_{P}=\Delta/\sqrt{2}\mu_B$. Taking into
account corrections for a possible $d$-wave symmetry ($B_{P}^{d}
\approx0.52\Delta/\mu_B$ \cite{Yang:98}), one obtains
$B_{P}\approx30-40$\,T using $\Delta\approx3.5$\,meV obtained
from our tunneling measurements. This value coincides well with
$B_{c2}^{\text{pg}}(T=0)$. However, one would then also expect
$B_{c2}^{\text{pg}}(T)\propto\Delta(T)$ what is not supported by
our data. In order to further clarify the experimental situation
as well as to bring more insight into the nature of the
superconducting and, in particular, the possible pseudogap state
in the electron doped HTS, more measurements at lower
temperatures, higher fields, and for different electron doping
levels, especially for the under(electron)doped case, are
required.

In conclusion, tunneling spectroscopy performed on bicrystal GBJs
has revealed a pseudogap feature in the low-energy quasiparticle
excitation spectrum of the electron doped HTS NCCO and PCCO. The
pseudogap feature evolves from the superconducting gap in applied
magnetic fields above the resistive critical field $B_{c2}^{\rho
}$ and vanishes at a four times higher field $B_{c2}^{\text{pg}}$
which is close to the Clogston limit. These experimental
observations suggest the existence of a pseudogap state also in
the electron doped HTS. The further clarification of the
pseudogap evidence for electron doped HTS is highly desired for a
more general understanding of the physical origin of the
pseudogap in the HTS.

This work is supported by the Deutsche Forschungsgemeinschaft (SFB
341).

\end{document}